# Asynchronous Early Output Section-Carry Based Carry Lookahead Adder with Alias Carry Logic

P. Balasubramanian, C. Dang, D.L. Maskell, and K. Prasad

*Abstract* - A new asynchronous early output section-carry based carry lookahead adder (SCBCLA) with alias carry output logic is presented in this paper. To evaluate the proposed SCBCLA with alias carry logic and to make a comparison with other CLAs, a 32-bit addition operation is considered. Compared to the weak-indication SCBCLA with alias logic, the proposed early output SCBCLA with alias logic reports a 13% reduction in area without any increases in latency and power dissipation. On the other hand, in comparison with the early output recursive CLA (RCLA), the proposed early output SCBCLA with alias logic reports a 16% reduction in latency while occupying almost the same area and dissipating almost the same average power. All the asynchronous CLAs are quasi-delay-insensitive designs which incorporate the delay-insensitive dual-rail data encoding and adhere to the 4-phase return-to-zero handshaking. The adders were realized and the simulations were performed based on a 32/28nm CMOS process.

## I. INTRODUCTION

Quasi-delay-insensitive (QDI) asynchronous circuits adopt an unbounded delay model for gates and wires with the exception of isochronic forks [1], which form the weakest compromise to delay-insensitivity. The signal transitions on all the isochronic forks, whether they are up-going or down-going, are assumed to happen concurrently. QDI circuits are the practically realizable delay-insensitive circuits which are robust to variations in process, supply and threshold voltages, and the operating temperature. Besides being adaptive and modular [2], QDI circuits are self-checking [3] and are naturally resistant to side channel attacks in the case of secure applications [4 – 7].

The main reasons for the robustness of QDI circuits are: i) delay-insensitive encoding for binary data representation and processing, and ii) adoption of a 4-phase handshake protocol for data communication. For delay-insensitive data encoding, the dual-rail or 1-of-2 code is widely used. The dual-rail or 1-of-2 code is the simplest member of the family of delay-insensitive *m*-of-*n* codes [8]. According to the dual-rail code, a data wire X is represented using two wires say, X1 and X0 as shown in Figure 1. X = 1 is encoded as X1 = 1 and X0 = 0, and X = 0 is encoded as X1 = 0 and X0 = 1. These two combinations represent the data. X1 = X0 = 0 is referred to as the spacer, and X1 = X0 = 1 is said to be invalid since the coding scheme is unordered. In this work, we consider the use of the dual-rail code for binary data encoding. The representation of 1 and 0 by respectively assigning a 1 to X1 and X0 on a mutually exclusive basis, and the usage of the zero spacer to denote the return-to-zero of all the encoded data wires defines the 4-phase return-to-zero handshake protocol [9]. According to the 4-phase return-to-zero protocol, the application of primary inputs to a QDI circuit follows the input sequence: data-spacer-data-spacer, and so forth.

P. Balasubramanian and C. Dang are with the School of Electrical and Electronic Engineering, Nanyang Technological University, 50 Nanyang Avenue, Singapore 639798, E-mails: balasubramanian@ntu.edu.sg; hcdang@ntu.edu.sg

D.L. Maskell is with the School of Computer Science and Engineering, Nanyang Technological University, 50 Nanyang Avenue, Singapore 639798, E-mail: asdouglas@ntu.edu.sg

K. Prasad is with the Department of Electrical and Electronic Engineering, Auckland University of Technology, Auckland 1142, New Zealand, E-mail: krishnamachar.prasad@aut.ac.nz

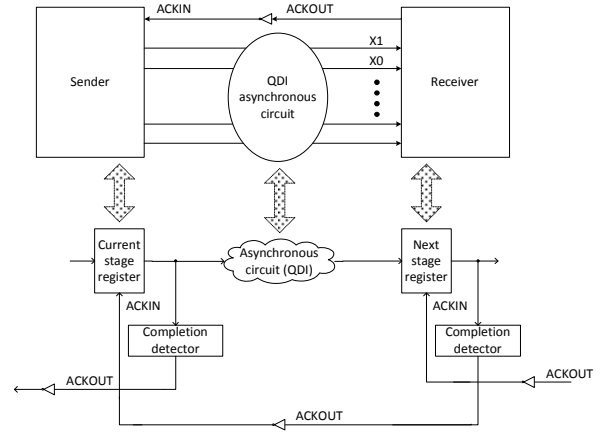

Fig. 1. A QDI asynchronous circuit stage correlated with the data sender and data receiver analogy for illustration

## II. QDI ASYNCHRONOUS CIRCUIT TYPES

QDI asynchronous logic circuits are classified as strongly indicating [10], weakly indicating [11] or early output type [12]. The input-output timing relation of strong-indication, weak-indication, and early output QDI asynchronous circuits is portrayed by Figure 2. A strong-indication asynchronous circuit will start to process the data or spacer to produce the corresponding primary outputs only after receiving all the primary inputs. A weak-indication asynchronous circuit could process the data or spacer after receiving just a subset of the primary inputs and may produce all but one of the corresponding primary outputs. However, a weak-indication asynchronous circuit would produce all the primary outputs only after receiving all the corresponding primary inputs whether they are data or spacer. An early output asynchronous circuit could process the data or spacer after receiving just a subset of the primary inputs and can produce all the corresponding primary outputs. Supposing an early output asynchronous circuit produces the spacer on all the primary outputs after receiving the spacer on only a subset of the primary inputs, it is said to be of early reset type. On the other hand, if an early output asynchronous circuit produces all the primary output data after receiving only a subset of the primary input data, it is said to be of early set type. The early set and reset behaviors of an early output asynchronous circuit are depicted through the portion encapsulated within the blue and red dotted ovals in Figure 2.

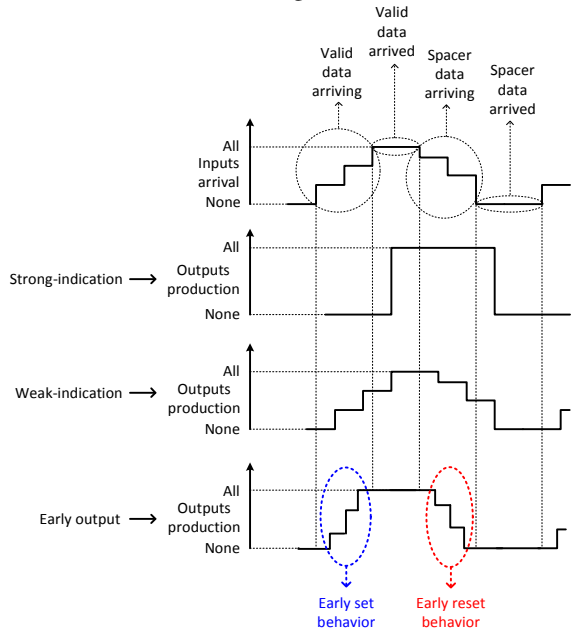

Fig. 2. Input-output timing relation of strong-indication, weak-indication and early output type QDI asynchronous circuits

It is important that a QDI asynchronous circuit should be devoid of circuit orphans viz. wire and gate orphans [13] [14]. Any unacknowledged signal transition on a wire is called wire orphan, and any unacknowledged signal transition on a gate output is called gate orphan. The signal transitions should be monotonic throughout the circuit i.e. monotonically increasing or monotonically decreasing in a QDI asynchronous circuit to ensure proper signal acknowledgment from the first logic level up to the last logic level [15] of the QDI circuit. Imposing the isochronic fork assumption on the primary inputs to a QDI asynchronous circuit would help to avoid the problem of wire orphan(s). This is because the completion detector that is present in each asynchronous circuit stage, as shown in Figure 1, will guarantee the complete arrival of the data and spacer into a QDI asynchronous circuit during the respective phases. Gate orphans are complicated to resolve than wire orphans and may necessitate imposing additional timing assumptions into a QDI circuit. Hence, the logic decomposition and physical synthesis of QDI asynchronous circuits have to be performed carefully by following safe QDI logic decomposition principles [16] [17]. In the next section, we present asynchronous early output SCBCLAs without and with the alias carry output logic.

## III. PROPOSED ASYNCHRONOUS EARLY OUTPUT SCBCLA WITHOUT/WITH ALIAS CARRY LOGIC

The SCBCLA is based on the concept of dividing an $n$-bit binary adder into $k$ sub-adder sections (i.e. $k$ sub-SCBCLA modules) where the size of each adder section is $m$-bits. Mathematically, $k = n/m$ where $k$, $m$ and $n$ are positive integers and are even. Here we consider $n = 32$ and $m = 4$. Hence a 32-bit SCBCLA is constructed using eight 4-bit sub-SCBCLA modules as shown in Figures 3a and 3b. In Figures 3a, 3b, 3f and 3g, (A311, A310) and (B311, B310) represent the most significant dual-rail augend and addend inputs, and (A01, A00) and (B01, B00) represent the least significant dual-rail augend and addend inputs. (C01, C00) denotes the dual-rail carry input and (C321, C320) denotes the dual-rail carry output. As seen in Figure 3, carry ripples within an adder section to produce the sum outputs, and the lookahead carry generated from an adder section is passed onto the next section as the carry input.

In Figures 3a, 3b, 3f and 3g, it can be seen that there is an inter-section propagation of the carry signal based on lookahead, and an intra-section propagation of the carry signal based on a simple rippling and both these tend to happen simultaneously. An SCBCLA uses the SCBCLG, the full adder (FA), and the sum only logic (SOL) as the circuit building blocks.

Figure 3c shows the gate-level detail of the 4-bit section-carry based carry lookahead generator (SCBCLG) without/with the alias carry output logic. The SCBCLG is different from a conventional CLG in that only one lookahead carry output is produced. If the circuit portion shown in red is removed from Figure 3c, then the 4-bit SCBCLG produces only the lookahead carry output (C41, C40). However, if the circuit portion shown in red is retained, then the 4-bit SCBCLG produces two pairs of lookahead carry outputs viz. (C41, C40) and (C41alias, C40alias). Note that these two dual-rail carry output pairs are logically equivalent. Figures 3d and 3e show the gate-level details of the early output FA and the early output SOL based on [12]. The SOL is identical to the FA but does not have a carry output.

The 4-bit SCBCLG shown in Figure 3c does not contain any redundant carry logic when it produces only the dual-rail carry output (C41, C40) and not the alias dual-rail carry output (C41alias, C40alias). However, when the alias dual-rail carry output is also produced by the 4-bit SCBCLG, then the 4-bit SCBCLG it is said to contain explicit logic redundancy [18].

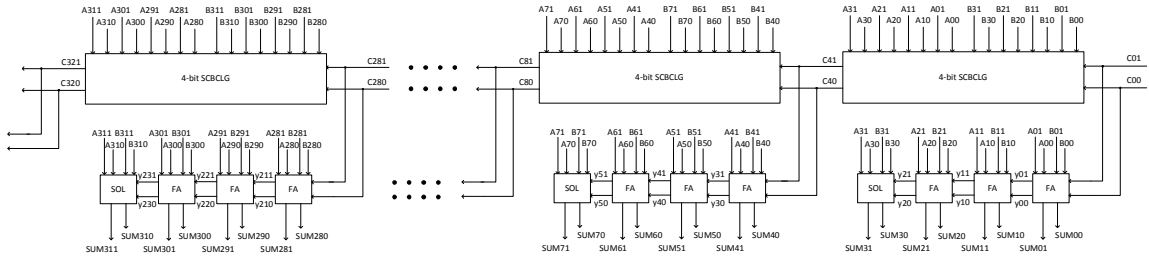

Fig. 3a. 32-bit early output SCBCLA without alias logic

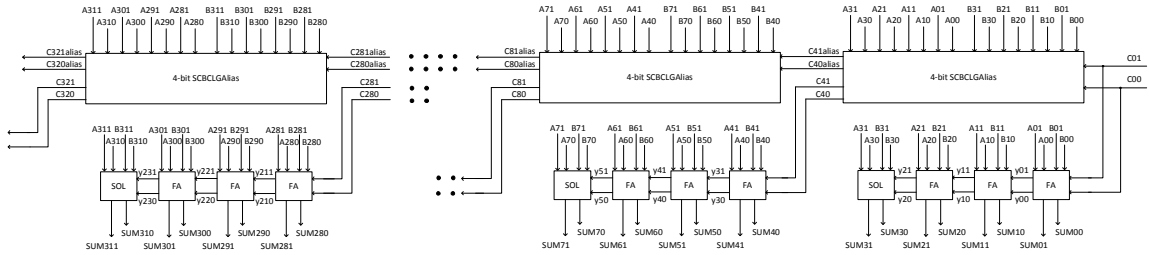

Fig. 3b. 32-bit early output SCBCLA with alias logic

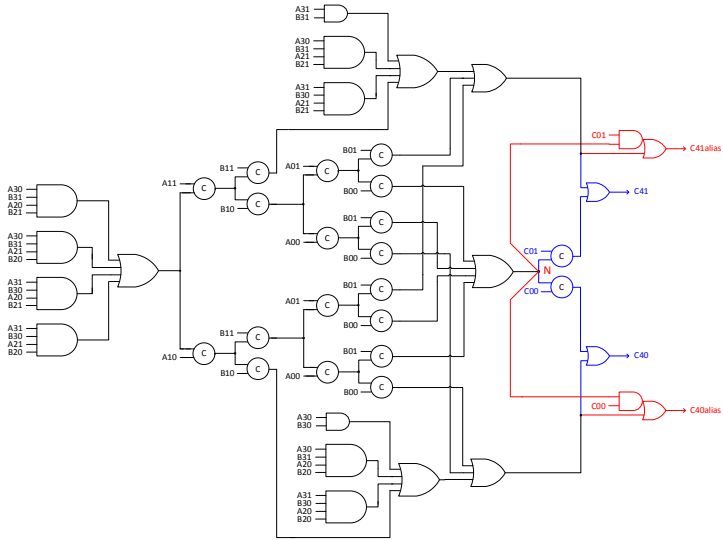

Fig. 3c. 4-bit SCBCLG without/with alias logic

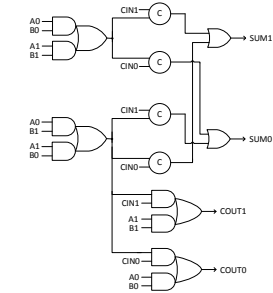

Fig. 3d. Early output full adder (FA)

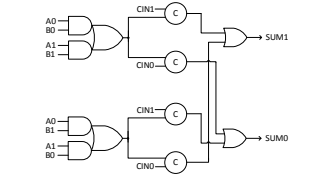

Fig. 3e. Early output sum only logic (SOL)

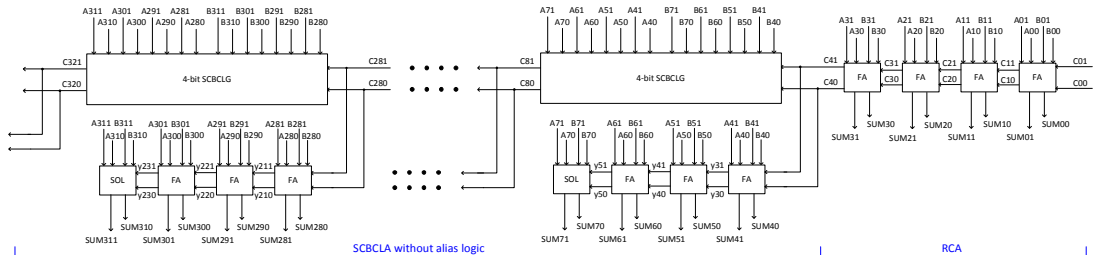

Fig. 3f. 32-bit early output SCBCLA (without alias logic) and RCA hybrid

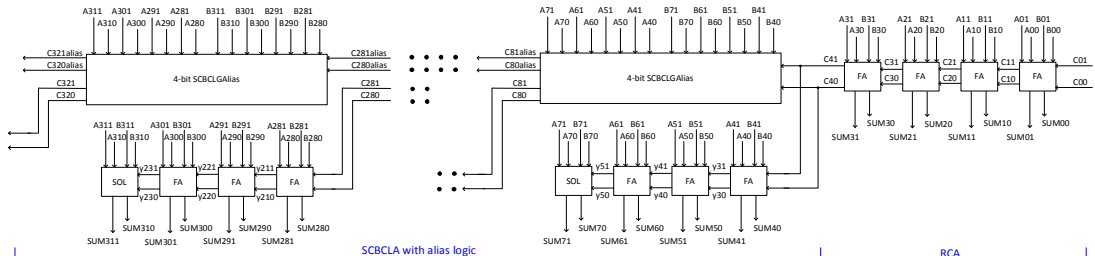

Fig. 3g. 32-bit early output SCBCLA (with alias logic) and RCA hybrid

The logic used to produce (C41, C40) is synthesized directly [19] based on deriving the disjoint sum-of-products form [20] [21] followed by QDI logic decomposition [17]. The carry output logic shown in blue in Figure 3e cannot be discarded and (C41alias, C40alias) cannot replace the dual-rail carry output (C41, C40) of the 4-bit SCBCLG due to the gate orphan problem. To explain this, let us assume that node N, shown in red in Figure 3c, and C01 are 1 during a data phase. In the following return-to-zero phase if C01 assumes 0 before N could assume 0, then C41alias may become 0 before N becomes 0. In this case, the late assumption of 0 by N would not be subsequently acknowledged by C41alias which results in a gate orphan. It may be noted at this juncture that when the 4-bit SCBCLG produces either (C41, C40) or (C41, C40) and (C41alias, C40alias), the presence of a C-element and an OR gate with respect to C41 and C40 eliminates the problem of gate orphan(s). For example, if C01 and N were 1 during a data phase, and if C01 assumes 0 before N assumes 0 then C41 would not become 0. This is because the C-element which has C01 and N as its inputs in Figure 3c would wait for the arrival of 0 on N. Only after N becomes 0, C41 would become 0. However, since node N is considered to be isochronic, the arrival of 0 on N would be deemed to be acknowledged by both C41 and C41alias.

When (C41, C40) is sufficient to serve as the lookahead carry output of the 4-bit SCBCLG shown in Figure 3c, what is the utility of the alias lookahead carry output (C41alias, C40alias)? In Figure 3c, it can be seen that between C01 and C41, a 2-input C-element[1] and a 2-input OR gate are present, which is the same with respect to C00 and C40. However, just a single complex gate viz. AO21 is used to connect C01 with C41alias, and likewise C00 with C40alias. If A, B, C and D are the inputs to an AO21 gate, the output of the AO21 gate, say $Y = AB + CD$. The AO21 gate requires just 10 transistors whereas the 2-input C-element and the 2-input OR gate combination requires 18 transistors. The propagation delay of an AO21 gate is less than the sum of the propagation delays of the 2-input C-element and the 2-input OR gate. Excepting the least significant 4-bit sub-SCBCLA, in the remainder of the sub-SCBCLAs, the inter-section carry propagation will be governed by the sum of the propagation delays of the 2-input C-element and the 2-input OR gate in the case of Figure 3a (32-bit SCBCLA without alias carry output logic). But in Figure 3b (32-bit SCBCLA with alias carry output logic), the inter-section carry propagation will be dictated by the propagation delay of just the AO21 gate. Hence, a faster inter-section carry propagation is feasible in the case of Figure 3b when compared to Figure 3a. This is an advantage of the alias carry output logic. Also, a faster return-to-zero can be facilitated by the alias carry output logic in the case of Figure 3b, while the return-to-zero in the case of Figure 3a would encounter the worst-case latency as for data processing.

Figures 3f and 3g portray two example hybrid 32-bit adders involving the SCBCLA and the ripple carry adder (RCA). Figure 3f shows an example 28-bit early output SCBCLA without alias logic which is combined together with a 4-bit early output RCA that is present in the least significant nibble position. Figure 3g shows an example 28-bit early output SCBCLA with alias logic that is joined to a less significant 4-bit early output RCA. The introduction of a RCA to replace the sub-SCBCLA or the sub-RCLA in the less significant adder positions was found to reduce the latency, area, and average power dissipation of a CLA and RCA hybrid in [19] and [22]. This shall be discussed in conjunction with the simulation results which are presented in the next section. The hybrid SCBCLA-RCAs shown in Figures 3f and 3g are just examples which are considered here to demonstrate their relative merits over the regular CLA counterparts.

## IV. SIMULATION RESULTS AND DISCUSSION

32-bit SCBCLAs without and with the alias carry output logic and 32-bit hybrid SCBCLA-RCAs without and with the alias carry output logic which correspond to weak-indication and early output types, and a 32-bit early output RCLA and a RCLA-RCA hybrid were all physically realized in semi-custom ASIC style using the standard cells of a 32/28nm CMOS process [23]. The 2-input C-element was alone custom designed using 12 transistors, and was made available to realize the various asynchronous CLAs. The size of the carry lookahead generator used in all the CLAs is 4-bits. For the hybrid SCBCLAs and RCLAs, a 4-bit least significant RCA was used. Only the minimum size cells of the standard digital cell library [23] were used for physical synthesis to enable a straightforward comparison between the synthesis results of different CLAs. Note that all the CLAs mentioned in Table I are QDI designs.

Approximately 1000 random input vectors were identically supplied to all the CLAs through a test bench at time intervals of 20ns to perform the functional simulations and also to capture their respective switching activities. The value change dump files generated through the functional simulations were then used for average power estimation using a Synopsys tool. The worst-case latency i.e. the critical path delay and the area occupancy of the CLAs were also estimated using the Synopsys tool. An appropriate wire load model (parasitic) was included while estimating the design metrics, which are given in Table I. The optimized design parameters are highlighted in bold-face in Table I. Since the input registers and completion detector of the various CLAs are identical, the differences between their design metrics is attributable to the differences between their respective function blocks.

The simulation results corresponding to various CLAs are split into four groups, labeled as Group1 to Group4 in Table I, for the sake of discussion. Group1 corresponds to regular and hybrid SCBCLAs without and with the alias carry output logic, which are weakly indicating. With respect to Group1, the 4-bit SCBCLG was realized based on a direct synthesis [19], and the FA and SOL are realized based on [24]. Since the FA and SOL of [23] are weakly indicating, the regular and hybrid SCBCLAs corresponding to Group1 also conform to weak-indication. Reference [25] presented a latency optimized weakly indicating FA design. This FA and the associated SOL were used to replace the FA and SOL components of the regular and hybrid SCBCLAs in Group 1, which yielded the Group2 results. Since the FA of [25] is more optimized compared to the FA of [24], therefore Group2 results are better compared to Group1 results as seen in Table I. The weak-indication FA of [24] occupies 41.17µm$^2$ of silicon, and the weak-indication FA of [25] occupies a reduced area of 39.65µm$^2$. The SOL based on [24] or [25] is the same and occupies 34.56µm$^2$ of silicon.

---
[1] The C-element outputs 1 if all its inputs are 1, and outputs 0 if all its inputs are 0. If its inputs are different, the C-element would maintain its existing steady-state. The C-element is symbolized by the circle with the marking 'C'.

TABLE I
AVERAGE POWER DISSIPATION, (WORST-CASE) LATENCY, AND AREA PARAMETERS OF VARIOUS 32-BIT ASYNCHRONOUS CLAS, ESTIMATED USING A 32/28NM CMOS PROCESS

| Results group | CLA or CLA-RCA hybrid adder type | Power (µW) | Latency (ns) | Area (µm$^2$) |
|---|---|---|---|---|
| *References* [19] [24]: *Weak-indication* | | | | |
| Group1 | SCBCLA (Without alias logic) | 2191 | 3.31 | 2951.88 |
| Group1 | SCBCLA-RCA hybrid (Without alias logic) | 2189 | 3.08 | 2845.14 |
| Group1 | SCBCLA (With alias logic) | 2192 | 2.46 | 2992.55 |
| Group1 | SCBCLA-RCA hybrid (With alias logic) | 2190 | 2.38 | 2880.72 |
| *References* [19] [25]: *Weak-indication* | | | | |
| Group2 | SCBCLA (Without alias logic) | 2188 | 3.14 | 2915.29 |
| Group2 | SCBCLA-RCA hybrid (Without alias logic) | 2186 | 2.93 | 2807.02 |
| Group2 | SCBCLA (With alias logic) | 2190 | 2.32 | 2955.95 |
| Group2 | SCBCLA-RCA hybrid (With alias logic) | 2187 | 2.25 | 2842.60 |
| *References* [22] [12]: *Early output* | | | | |
| Group3 | RCLA | 2177 | 2.75 | 2569.65 |
| Group3 | RCLA-RCA hybrid | **2175** | 2.53 | 2455.80 |
| *Proposed: Early output* | | | | |
| Group4 | SCBCLA (Without alias logic) | 2178 | 3.13 | 2524.92 |
| Group4 | SCBCLA-RCA hybrid (Without alias logic) | **2175** | 2.92 | **2416.66** |
| Group4 | SCBCLA (With alias logic) | 2179 | 2.31 | 2565.58 |
| Group4 | SCBCLA-RCA hybrid (With alias logic) | 2177 | **2.23** | 2452.24 |

Group3 comprises a regular RCLA based on [22] and a RCLA-RCA hybrid based on [22] and [12]. The 4-bit RCA employed in the RCLA-RCA hybrid is composed of four FA modules, and the FA is based on [12]. Since the sub-RCLAs and the RCA are early output type, therefore the RCLA and the RCLA-RCA hybrid also correspond to early output type. There is no possibility for introducing an alias carry output logic in the case of the RCLA or the RCLA-RCA hybrid. This is because the lookahead carry output of one sub-RCLA directly serves as the carry input for the successive sub-RCLA. In the regular or hybrid SCBCLAs, however, the lookahead carry output generated from one sub-SCBCLA serves as the carry input for the next SCBCLG and also as the carry input for the sub-RCA embedded within the successive sub-SCBCLA. Due to the supply of two dual-rail carry inputs to a sub-SCBCLA, the alias carry output logic was able to be introduced which enables significant optimization in the latency at the expense of just meagre increases in area and average power due to the redundant alias carry logic.

Group4 comprises the proposed regular and hybrid SCBCLAs without/with the alias carry output logic, which corresponds to early output type. This results from the use of the early output 4-bit SCBCLG without/with the alias carry output logic, the early output FA, and the early output SOL. The early output plain 4-bit SCBCLG requires 113.35µm$^2$ of silicon, and the silicon requirement increases to 118.43µm$^2$ with the introduction of the alias carry output logic. The early output FA and SOL require reduced areas compared to the weak-indication FA and SOL of [25] of just 27.45µm$^2$ and 22.36µm$^2$ of silicon respectively. Since the early output asynchronous circuits are more relaxed compared to their strong- and weak-indication circuit counterparts, simple and complex logic gates of a digital cell library can be widely used compared to the C-element. As a result, early output asynchronous circuits generally facilitate enhanced optimizations in the design metrics compared to the strong- and weak-indication asynchronous circuits. This is the primary reason for the Group4 results being more optimized compared to the synthesis results of Group1, Group2 and Group3.

Three important observations can be made from Table I. Firstly, the SCBCLAs with alias carry logic report a substantial reduction in latency compared to the SCBCLAs without the alias carry logic, and due to the redundant logic introduced in the case of the former their area and power metrics are marginally more expensive compared to the latter. On average, the SCBCLAs with alias carry output logic which correspond to Group1, Group2 and Group4, whether they are regular or hybrid variants, report a 24.6% reduction in latency and a 1.4% increase in cells area with negligible power increase (0.1%) compared to the averaged design metrics of the regular and hybrid SCBCLAs which have no alias carry output logic. This implies the SCBCLAs featuring the alias carry output logic achieve a significant reduction in latency with almost no increase in the area and power dissipation.

Secondly, the SCBCLA-RCAs hybrid and the RCLA-RCA hybrid enable additional optimizations in the design metrics compared to the regular SCBCLAs and RCLA. On average, the SCBCLA-RCAs hybrid without alias logic and the RCLA-RCA hybrid report a 7% reduction in latency and a 4% reduction in area with no power increase compared to the regular SCBCLAs without alias logic and the regular RCLA. Likewise, the SCBCLA-RCAs hybrid with alias carry logic, on average, report a 3% reduction in latency and a 4% reduction in area without any power increase compared to the regular SCBCLAs with alias carry logic. The area reduction is because a sub-SCBCLA without/with the alias carry logic and a sub-RCLA are more area expensive than a similar size RCA. Further, the critical path delay of the least significant 4-bit SCBCLA with alias carry logic (Figure 3b) is governed by the sum of the propagation delays of a 4-input AND gate, two 4-input OR gates, four 2-input C-elements and an AO21 gate. On the other hand, the least significant 4-bit RCA shown in Figure 3g encounters the sum of the propagation delays of five AO22 gates. Hence, Figures 3f and 3g will exhibit reduced latencies than Figures 3a and 3b. Thus using a small optimum size RCA to replace the sub-SCBCLAs or the sub-RCLA in the least significant positions is beneficial for reducing the area, latency and power parameters, which is substantiated by the results given in Table I.

Thirdly, it is clear from Table I that the proposed 32-bit SCBCLA-RCA hybrid incorporating the alias carry output logic features the least latency and is preferable. It is slightly more expensive in area than the 32-bit SCBCLA-RCA hybrid with no alias carry output logic by just 1.5% and the power increase is negligible (0.1%).

## V. CONCLUSIONS

This paper has presented a new asynchronous early output SCBCLA architecture without/with the alias carry output logic. The 32-bit binary addition was considered as the case study and the proposed SCBCLA with alias carry output logic reports improvements in design metrics than the other SCBCLAs and RCLAs proposed earlier.